\documentclass[ASNA,twocolumn]{USG} 
\usepackage{anyfontsize} %

\usepackage{colortbl}   
\usepackage{xcolor}     

\usepackage{steinmetz}
\graphicspath{{./images/}}

\articletype{Research Article}%
\subarticletype{Research Article}

\journal{Journal of Computer Assisted Learning}
\volume{0}
\copyyear{2026}
\startpage{1}
\articledoi{}

\begin{document}
\title{Micro-Level AI Feedback Features and Student Responses in Consecutive LLM Tutoring Interactions}
\author{Shayla Sharmin}[https://orcid.org/0000-0001-5137-1301]
\author{Mohammad Fahim Abrar}[https://orcid.org/0009-0009-5157-7807]
\author{Mohammad Al-Ratrout}[https://orcid.org/0009-0002-0180-4345]
\author{Roghayeh Leila Barmaki}[https://orcid.org/0000-0002-7570-5270]

%
\authormark{Anonymous \textsc{et al.}}
\titlemark{Micro-Level AI Feedback Features and Student Responses in Consecutive LLM Tutoring Interactions}

\address[1]{\orgdiv{Department of Computer and Information Sciences, }\orgname{University of Delaware, }\orgaddress{\state{DE, }\country{USA}}}

\corres{Shayla Sharmin  (\email{shayla@udel.edu})}


\presentaddress{This is sample for present address text. This is sample for present address text.}

\renewcommand{\sectionautorefname}{Section}
\renewcommand{\subsectionautorefname}{Section}
\renewcommand{\subsubsectionautorefname}{Section}


\abstract[ABSTRACT]{
\textbf{Background:} AI-assisted feedback can vary in explanatory content, emotional support, and response length, yet most prior research evaluates its effects at the task or session level. Less is known about how specific features of an AI response relate to students’ observable states in the immediately following interaction, particularly after students express confusion.

\textbf{Objectives:} This study examined whether concrete elaboration, affective language, and AI response length were associated with re-confusion and observed understanding in students’ subsequent interactions with an LLM tutor.

\textbf{Methods:} We conducted a secondary analysis of 16,851 user–AI interactions from the StudyChat dataset, collected from students using an LLM-based tutor in an undergraduate artificial intelligence course. An LLM-based annotation pipeline identified 1,718 confusion windows from 152 users, and logistic Generalized Estimating Equation models examined associations between AI-response characteristics and next-turn outcomes while accounting for repeated observations within users.

\textbf{Results and Conclusions:} Concrete elaboration, including analogies, comparisons, and examples, was associated with a greater likelihood of observed understanding in the subsequent interaction, whereas affective language showed no detectable association with either outcome. Longer AI responses were associated with lower observed understanding and greater re-confusion. These findings suggest that turn-level analysis can reveal distinct patterns that are obscured in session-level evaluations and motivate controlled studies testing whether concise, targeted elaboration can better support students following confusion.
}

 \transabstract[transABSTRACT]{This is a generic template designed for use by multiple journals, which includes several options for customization. Please refer to the author guidelines and author LaTeX manuscript preparation document for the journal to which you
are submitting in order to confirm that your manuscript will
comply with the journal’s requirements. Please replace this
text with your abstract. This is sample abstract text just for the template display purpose.}

\abbr{}


\copyright{This is an article under the terms of the \href{Creative Commons Attribution-NonCommercial}{Creative Commons Attribution-NonCommercial} License, which permits use, distribution and reproduction in any medium, provided the original~work~is~properly cited and is not used for commercial purposes.}


\maketitle


\section*{Practitioner Notes}
\begin{itemize}

\item[]\textbf{What is already known about this topic}
    \item AI-assisted feedback can support learning and student experience, but most studies examine task- or session-level outcomes.
    \item Concrete explanations, encouraging language, and response length may influence how students respond to AI feedback.
    \item Less is known about students' immediate responses after they express confusion.

\item[]\textbf{What this paper adds}
    \item This study examines AI feedback across consecutive interactions after student confusion.
    \item Concrete elaboration, such as analogies, comparisons, and worked examples, was associated with greater observed understanding in the next interaction.
    \item Affective language showed no clear association with next-turn outcomes, while longer responses were associated with lower observed understanding and greater re-confusion.

\item[]\textbf{Implications for practice and/or policy}
    \item AI tutoring systems should consider both the content and length of feedback.
    \item Focused explanations with examples or comparisons may be preferable to simply providing longer responses when students are confused.
    \item Controlled studies are needed before these findings are used as firm design recommendations.

\end{itemize}
\section{Introduction}\label{sec:introduction}
As AI-powered learning tools become increasingly common, AI-assisted feedback (AIFB) systems have emerged as a scalable way to help students recognize misunderstandings and improve their thinking \citep{BA2025Review}. Along with providing correct information, how it is delivered is also important. Micro-level features, such as concrete elaboration, affective language, and response length, represent different ways in which AI feedback may support learning.

Concrete elaboration can connect new concepts to existing knowledge by grounding an explanation in something tangible rather than simply correcting an error. Concrete elaboration includes both comparison-based explanations, sometimes called analogies (e.g., relating a concept to a familiar idea), and worked examples (e.g., illustrating a concept with a small, runnable code snippet that shows the concept in action). Prior work has examined the use of LLM-generated analogies to support understanding and learning \citep{Bernstein2024AnalogyPref,Shao2025AnalogyLLM,CAO2024AnalogyNovice}.

Affective language, which acknowledges student difficulty or offers encouragement, serves a different purpose. It focuses on the learner's emotional experience by recognizing frustration, uncertainty, or effort. Prior research has explored emotionally enriched and empathic AI-generated feedback in educational interactions \citep{alsaiari2024emotionally,Ortega2024EmathyChatbot,siyan2024eden}.

\begin{figure*}[h]
    \centering
    \includegraphics[width=\linewidth]{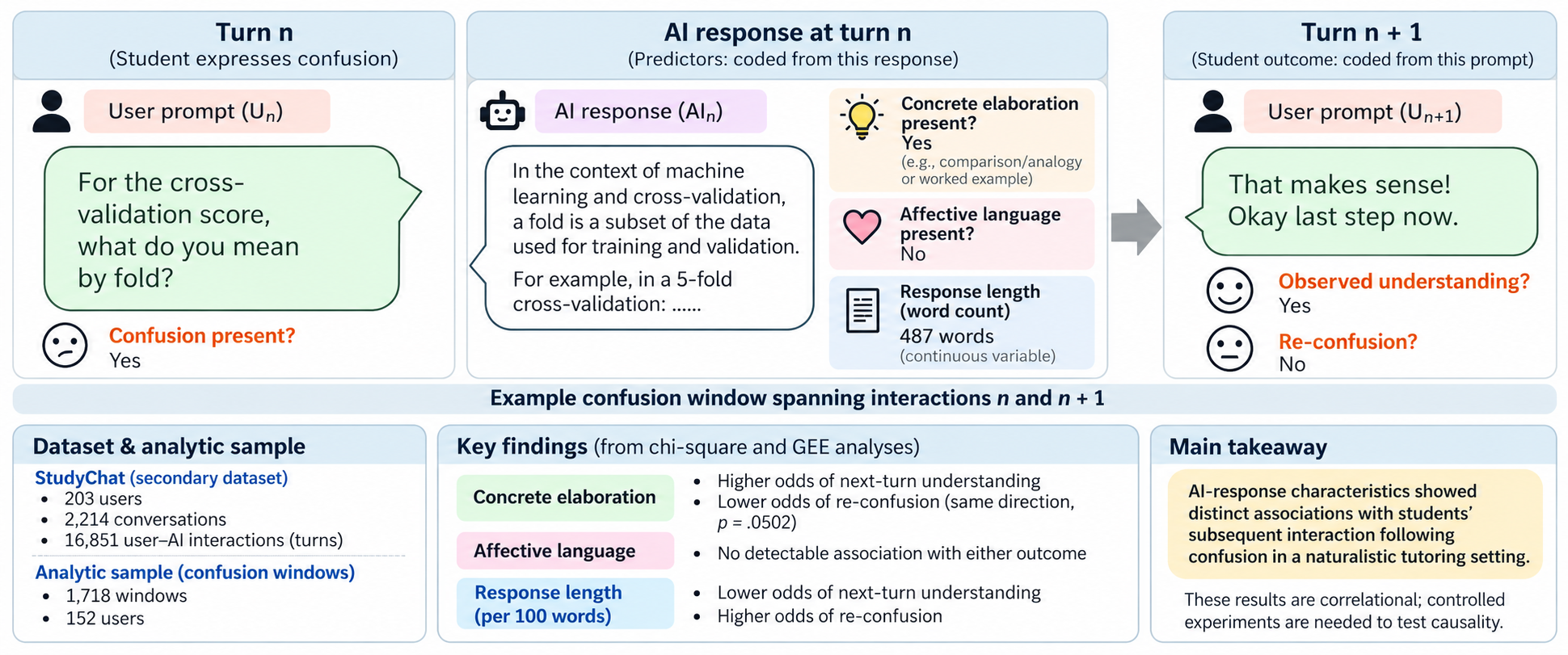}
    \caption{Illustration of a confusion window from the StudyChat dataset. 
    A confusion window consists of two consecutive user--AI interactions 
    (\(U_n\)--\(AI_n\) and \(U_{n+1}\)--\(AI_{n+1}\)). In this example, 
    \(AI_n\) contains the micro-level feedback feature concrete elaboration 
    (i.e., a worked code example used to clarify a concept), while 
    \(AI_{n+1}\) illustrates the feature affective language. For the analysis, 
    all predictors (concrete elaboration, affective language, and response length) 
    are coded from the AI response at \(AI_n\), whereas both outcomes 
    (understanding and re-confusion) are coded from the student's subsequent 
    prompt at \(U_{n+1}\). \(AI_{n+1}\) is shown only to illustrate the affective 
    language label; affective language was coded from \(AI_n\) in the actual analysis.}
    \label{fig:confusion-window}
\end{figure*}

Response length may also matter; longer explanations can provide additional information but may also increase the amount of information students must process \citep{shutte2008feedback,sweller1988cognitive,paas2003cognitive}. Most AIFB studies evaluate outcomes at the task or session level. They show whether students learned after completing an activity. In this study, we focused on what happens during the interaction itself by examining consecutive user--AI interactions (\autoref{fig:confusion-window}).

A consecutive user--AI perspective offers an opportunity to examine how students respond immediately after receiving different kinds of AI feedback. It allows us to explore whether features such as concrete elaboration, affective language, and response length are associated with students' observable states from one interaction to the next. We used the StudyChat dataset to examine these micro-level AIFB features in consecutive interactions \citep{mcnichols2026studychat}. It contains 16,851 conversational rows that consist of user prompts and AI feedback from real student interactions with a large language model (LLM) tutor in an undergraduate AI course.

Specifically, we identified confusion windows in which a student expressed confusion at interaction $n$, characterized the AI response at that interaction, and examined re-confusion and observed understanding in the student's subsequent interaction ($n+1$). We focus on consecutive interactions to answer the following three questions:

\begin{itemize}
    \item \textbf{RQ1:} To what extent is concrete elaboration in an AI response associated with re-confusion and observed understanding in the student's subsequent interaction?
    
    \item \textbf{RQ2:} To what extent is affective language in an AI response associated with re-confusion and observed understanding in the student's subsequent interaction?
    
    \item \textbf{RQ3:} To what extent is AI response length associated with re-confusion and observed understanding in the student's subsequent interaction, after accounting for concrete elaboration and affective language?
\end{itemize}

Most research on AI feedback evaluates outcomes after a task or session, making it difficult to examine what happens immediately after a student expresses confusion. This study instead uses the consecutive interaction following a confused student turn as the unit of analysis. Using an LLM-based annotation pipeline applied to 16,851 naturalistic tutoring interactions, we identify 1,718 confusion windows and examine whether concrete elaboration, affective language, and AI response length are associated with re-confusion and observed understanding in the subsequent interaction. This turn-level perspective complements session-level analyses by distinguishing response characteristics that may be associated with students' immediate observable states following confusion.

\section{Related work} \label{sec:related}
\subsection{Feedback frameworks and the role of elaborated feedback in AIFB}

In education, micro-level feedback concerns characteristics of how feedback is presented within individual feedback instances, including its level of elaboration \citep{BA2025Review}. In AI-assisted feedback, features such as concrete elaboration, affective language, and response length provide distinct ways of characterizing the content, tone, and amount of information in an individual response. More broadly, LLMs can take different roles in educational settings, from supporting understanding and practice to directly generating outputs for learners \citep{shayla2026youtube}. AI-assisted feedback has been associated with improvements in cognitive outcomes, satisfaction, motivation, self-confidence, and positive emotions in learning \citep{BA2025Review}. Students perceived AI feedback as highly reliable, and using the system more to solve more exercises was also associated with better exam performance \citep{Letteri2025EnhanceFeedback}. Students also tend to take up meaning-level corrections more than surface-level corrections, such as grammar fixes, and instead of trusting AI suggestions blindly, they use strategies like internet searches and proofreading to check the suggestions \citep{Xu2025EngagementAIfeedback}. In writing, feedback generated by GPT led to clear improvement in revising text inside a real writing class, along with more motivation and positive feelings, though it did not help students do better on a new writing task afterward \citep{MEYER2024LLMFeedbackClassroom}. In online discussions, idea-focused ChatGPT feedback was found to significantly increase student engagement and promote deeper thinking compared to task-focused feedback in a real higher education course \citep{Ba2025ImpactAIFeedback}. AI feedback with greater warmth has also been associated with students’ emotional experience during learning \citep{alsaiari2024emotionally}.

 \subsection{Micro-level feedback features: concrete elaboration, affective language, and response length}

Concrete elaborations, including analogies and worked examples, can connect new concepts to learners’ prior knowledge and may support understanding by making abstract information more concrete \citep{paas2003cognitive,sweller1988cognitive}. LLMs now generate such analogies in real time, tailored to individual learners \citep{jurenka2024towards}, and personalized analogies are associated with better understanding, confidence, memory, efficiency, and overall learning experience compared to generic ones \citep{CAO2024AnalogyNovice}. However, students also want complementary supports such as examples, visualizations, and technical explanations, suggesting that analogies work best alongside other instructional support \citep{CAO2024AnalogyNovice}. While survey studies show students generally perceive AI-generated analogies as accurate, interesting, and useful \citep{Bernstein2024AnalogyPref}, these findings reflect perception rather than learning outcomes. Classroom evidence is more nuanced: analogy usefulness depends on instructional context and the quality of existing materials, with students perceiving less need for analogies when textbook explanations are already clear \citep{Shao2025AnalogyLLM}. Still, improved AI-generated analogies are linked to better homework performance, increased classroom attention, and inspiration for teachers’ own explanations. Researchers caution, however, that without teacher guidance, students may over-rely on the analogy itself rather than the underlying material, risking superficial understanding. Beyond analogies, elaborated AI feedback shows similar benefits in writing: AI-generated explanations improve writing outcomes, and students often find combined human-AI feedback more useful than either alone \citep{Escalante2023Writing}. These studies, though, measure outcomes at the session or draft level rather than the effect of a single feedback instance. Such feedback need not be delivered only after task completion. Recent experimental work further suggests that real-time AI support delivered during learning—including explanations, analogies, comparisons, and Socratic prompting—can support subsequent performance \citep{google2025Feedback,wang2024tutor}.

Prior studies suggest that affective AI feedback may influence students’ emotional experiences more consistently than behavioral or learning outcomes. Students find feedback enriched with praise, encouragement, emojis, and motivational language more helpful because it significantly reduces anger and demotivation compared to neutral-tone feedback. However, it does not necessarily affect engagement or the quality of the work produced \citep{alsaiari2024emotionally}. Brain signal data analysis also shows that affective language (praise/encouragement) helps with shallow retention, but its effect on deeper learning (transfer, metacognition) is similar to that of a neutral feedback system \citep{Yin2025brainActivity}. Feedback combining cognitive and affective elements shows a similar pattern. Although affective language correlated with "conceptual change" and "personal growth," group-level changes in learning performance, motivation, or self-regulation were limited \citep{Ortega2024EmathyChatbot}. Some studies focus on perception rather than behaviour. For example, \citet{siyan2024eden} compared none/fixed/adaptive affective language mainly on perceived affective support and self-reported grit, without measuring performance. Similarly, \citet{kar2025mathbuddy} compare Emotion ON vs Emotion OFF and establish a framework linking learning gains to engagement and perceived support.

Response length is a potentially important and controllable design characteristic in AI-supported learning, as the amount of information presented may shape how students engage with and process feedback. In emotionally enriched feedback settings, longer responses, often including praise, emojis, and conversational framing, can improve students’ affective experience, but they may also reduce the likelihood that students read the full message when compared to more concise feedback \citep{alsaiari2024emotionally}. Consistent with this, several studies report that learners tend to prefer brief feedback, typically only a few sentences, with relatively low preference for extended responses. Reflecting this preference, some systems explicitly operationalized length constraints in their design, such as offering selectable “succinct” versus “detailed” modes \citep{siyan2024eden}, standardizing feedback length across conditions to isolate content effects \citep{Yin2025brainActivity}, or limiting the number of interactions per turn to avoid cognitive overload \citep{kar2025mathbuddy}. From the perspective of Cognitive Load Theory, working-memory capacity is limited, and excessive or non-essential information can increase extraneous processing demands, potentially making task-relevant information more difficult to identify and process \citep{paas2003cognitive,sweller1988cognitive}. This concern may be particularly relevant in moments of learner confusion, when attention should be directed toward resolving specific knowledge gaps rather than processing extended explanations. However, despite its practical and theoretical relevance, empirical work directly examining associations between AI response length and immediate subsequent-turn outcomes in tutoring interactions remains limited.

\subsection{Confusion resolution in conversational learning}

Recent work on confusion in LLM-based tutoring dialogues approaches the problem from several angles. \citet{liu2025discerning} introduce a benchmark from real educational dialogues with controlled student states, showing that current LLM tutors often struggle to adapt when a student is confused, responding with vague comments or topic shifts rather than targeted scaffolding; however, response quality here is judged holistically rather than by specific features like tone, length, or concrete elaboration such as analogy use. Other work takes a conversation-level view. In the tutoring dataset proposed by \citet{macina2023mathdial}, teachers resolved confusion in about 89\% of conversations, mainly through scaffolding questions, yet the interaction right after confusion is never isolated or linked to specific tutor behaviours. A shared task by \citet{kochmar2025findings} moves closer to a turn-level view, evaluating individual tutor responses to student mistakes or confusion along dimensions like mistake identification and guidance versus direct answers. Without access to the student’s next interaction, though, it cannot assess whether confusion was resolved. \citet{Alexandar2025LLM}, meanwhile, predict whether a tutor response will lead a student to answer correctly on the next turn, using features like hinting versus revealing answers. Here, too, the outcome captured is next-turn correctness rather than confusion resolution specifically, and confusion is never isolated as the triggering condition.

\subsection{Scope}

Taken together, prior work on AI-supported feedback and tutoring highlights three complementary strands: (a) studies of response-level characteristics such as concrete elaboration, affective language, and response length; (b) research on conversational tutoring strategies for responding to student confusion; and (c) work examining subsequent student outcomes following tutor responses. Although these studies provide valuable insights, they commonly focus on aggregated task- or session-level outcomes, evaluations of individual tutor responses without observing the subsequent student turn, or next-turn correctness without isolating confusion as the triggering condition. Consequently, comparatively little is known about how specific characteristics of an AI response are associated with a student’s observable state immediately after expressing confusion. In particular, it remains unclear whether concrete elaboration, affective language, and response length show different associations with subsequent re-confusion and observed understanding. This study addresses this gap by examining these three response characteristics within consecutive interaction windows drawn from naturalistic LLM tutoring conversations.
\section{Methods}
\label{sec:methods}

\subsection{Dataset}
We used the StudyChat dataset, a publicly available dataset on HuggingFace (wmcnicho/StudyChat) containing conversation logs from an LLM-based tutoring assistant (GPT-4o mini) deployed in an undergraduate artificial intelligence course \citep{mcnichols2026studychat}.
Students used it for open-ended programming help 
across seven assignments during the Fall 2024 and Spring 2025 semesters. The final 
dataset contained 16,851 user--AI interactions from 203 unique users across 2,214 
conversations. We sorted all turns by \texttt{chatId} and 
\texttt{interactionCount} to reconstruct the sequence of events within each 
conversation. The original dataset creators handled ethical approval, informed 
consent, and anonymization. We performed secondary analysis only on this publicly 
released dataset.

\subsection{LLM-based Annotation Pipeline}

We used a locally hosted open-weight model (Gemma 4, 26B parameters; Ollama model 
identifier: 
\texttt{gemma4:26b}), served via Ollama, to 
annotate each user--AI interaction consecutively in the dataset.
Each row, or 
interaction \(n\), for every \texttt{chatID} in the dataset contains a user prompt 
(\(U_n\)) and AI-assisted feedback (\(AI_n\)). Each interaction received binary 
labels (0 or 1) across six dimensions defined in \autoref{tab:annotation-labels}.

Each annotation call allowed up to three retries. When all retries failed, we 
assigned conservative default labels (all values \(= 0\)), and checkpoints were 
saved every 100 rows to allow recovery from interruptions. The reliability of the LLM-based annotations was assessed against independent 
human coding (\autoref{sec:human-validation}).

\begin{table*}[t]
\centering
\caption{Label definitions used in the annotation pipeline.}
\label{tab:annotation-labels}
\begin{tabularx}{\linewidth}{
    >{\raggedright\arraybackslash}p{0.31\linewidth}
    >{\raggedright\arraybackslash}X
}
\toprule
\textbf{Label} & \textbf{Definition} \\
\midrule

\texttt{is\_\(U_n\)\_new\_context}
&
1 if \(U_n\) is the first interaction of the specific chat or a new topic. 
Always the binary complement of \texttt{is\_\(U_n\)\_followup}. 
\\

\addlinespace

\texttt{is\_\(U_n\)\_followup}
&
1 if \(U_n\) continued the same topic as \(U_{n-1}\); 0 if the student 
switched to a new topic. Always the binary complement of 
\texttt{is\_\(U_n\)\_new\_context}. 
\\

\addlinespace

\texttt{confusion\_in\_\(U_n\)}
&
1 if the student repeated the question, rephrased the same confusion, or 
showed the same error as \(U_{n-1}\). Only possible when 
\texttt{is\_\(U_n\)\_followup} \(= 1\). This label marks the current interaction 
as confused---it is not an outcome variable. 
\\

\addlinespace

\texttt{understanding\_in\_\(U_n\)}
&
1 if the student showed explicit understanding (e.g., ``I see,'' 
``that makes sense,'' or correct application of the concept) in \(U_n\), 
indicating understanding of the preceding AI response \(AI_{n-1}\). 
\\

\addlinespace

\texttt{elaboration\_present\_\(AI_n\)}
&
1 if the AI response used a concrete elaboration such as an analogy, 
comparison, or example to explain a concept. 
\\

\addlinespace

\texttt{affective\_language\_present\_\(AI_n\)}
&
1 if the AI response acknowledged student difficulty or offered encouragement. 
\\

\bottomrule
\end{tabularx}
\end{table*}

The model annotated the current user prompt \(U_n\), indicating whether it showed 
confusion or understanding and checking whether it was a follow-up to the previous 
interaction (\(U_{n-1}\)). For the current AI response (\(AI_n\)), the model 
annotated the presence of concrete elaboration and affective language.

\noindent\textbf{Response length:}
We computed the word count of each AI response and retained response length as a continuous measure. For regression analyses, word count was scaled in units of 100 words to provide interpretable effect estimates. Thus, the corresponding odds ratio represents the change in the odds of the outcome associated with an additional 100 words in the AI response.

\noindent\textbf{Interaction-level annotation procedure:}
Because the interpretation of a user prompt depends on its position within a conversation, we distinguished between the first interaction and subsequent interactions during annotation. The first interaction has no preceding user–AI exchange, whereas subsequent interactions can be evaluated in relation to the preceding exchange. We therefore applied different annotation procedures depending on the interaction's position within each conversation, as described below.
\begin{itemize}
    \item \noindent\textbf{First-interaction labelling:}
In the first interaction of each conversation, the user prompt (\(U_n\)) was treated as a new context with no confusion or understanding expressed by the user (\(\mathrm{confusion\_in\_}U_n = 0\), \(\mathrm{understanding\_in\_}U_n = 0\), \(\mathrm{is\_}U_n\mathrm{\_new\_context} = 1\)). Because no prior AI response existed, only the AI response (\(AI_n\)) was labelled for concrete elaboration and affective language.
\item \noindent\textbf{Subsequent-interaction labelling:}
For all later interactions, the model received the previous user--AI interaction (\(U_{n-1}\)--\(AI_{n-1}\)), the current user prompt (\(U_n\)), and the current AI response (\(AI_n\)). It determined whether \(U_n\) continued the previous topic or introduced a new context and, for follow-up turns, whether \(U_n\) expressed confusion or understanding of \(AI_{n-1}\). The current AI response (\(AI_n\)) was also labelled for concrete elaboration and affective language.
\end{itemize}

\noindent\textbf{Annotation consistency rules:}
We enforced three logical consistency rules after each annotation call. First, \(\mathrm{is\_}U_n\mathrm{\_new\_context}\) was always set as the complement of \(\mathrm{is\_}U_n\mathrm{\_followup}\). Second, \(\mathrm{confusion\_in\_}U_n\) was set to 0 whenever \(\mathrm{is\_}U_n\mathrm{\_followup} = 0\), as confusion was only annotated in follow-up turns. Third, confusion and understanding were mutually exclusive; when both were flagged, confusion was retained, and understanding was set to 0. Understanding could still be coded when a student moved to a new topic after explicitly confirming comprehension (e.g., ``Got it, thanks---now for the next part\ldots'').

\subsection{Variables and Measures}

Windows were constructed by pairing consecutive interactions 
(\(U_n\)--\(AI_n\) and \(U_{n+1}\)--\(AI_{n+1}\)) within each conversation, 
reusing labels already produced by the annotation pipeline. The unit of analysis 
was a confusion window: a pair where the student showed confusion at interaction 
\(n\), the AI response at \(n\) provided the predictor variables, and the student's 
response at \(n+1\) provided the outcome. Because \texttt{confusion\_in\_\(U_n\)} could only be coded for follow-up turns, a confusion window captures persisting confusion about the preceding AI response rather than a student's first expression of difficulty. When multiple confused turns occurred 
consecutively, we created overlapping windows; the statistical models therefore 
accounted for repeated observations within users.

\begin{figure*}[h]
    \centering
    \includegraphics[width=\linewidth]{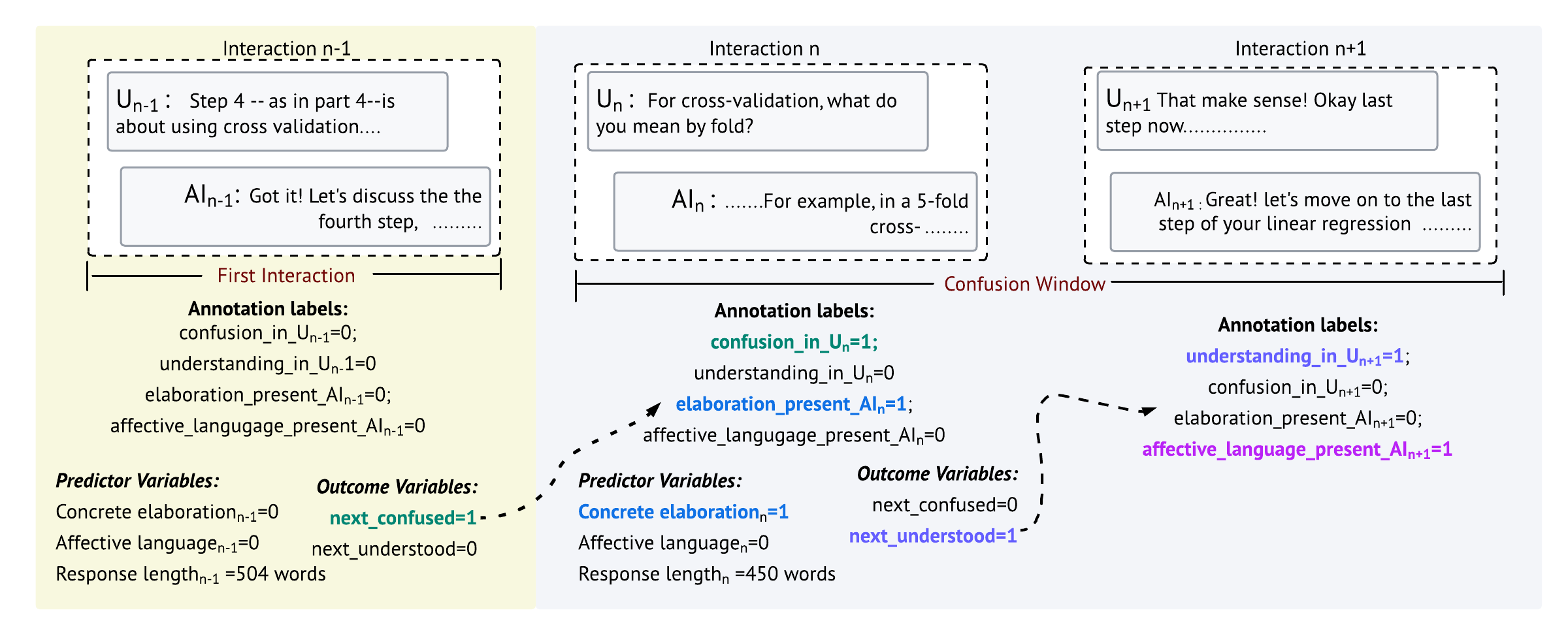}
    \caption{Annotation labels and window construction across three consecutive 
    user--AI interactions. Student-level labels 
    (\(\mathrm{confusion\_in\_}U_n\), 
    \(\mathrm{understanding\_in\_}U_n\)) are assigned by relating \(U_n\) 
    to the preceding AI response (\(AI_{n-1}\)) (dashed arrow); AI-level labels 
    (\texttt{elaboration\_present}, \texttt{affective\_present}) are assigned 
    to the current AI response. Response length is computed directly from the 
    word count of each AI response.}
    \label{fig:annotation-window}
\end{figure*}

Of 16,851 total interactions, 1,912 were labelled 
\(\mathrm{confusion\_in\_}U_n = 1\); after excluding 194 terminal turns 
(no subsequent interaction available), 1,718 confusion windows remained, drawn 
from 654 conversations and 152 users. Each window links an AI response at 
interaction \(n\) to the student's subsequent response at \(n+1\), allowing us 
to examine associations between characteristics of the AI response and the 
student's observed state in the immediately subsequent turn. From the six labels, 
we constructed three predictor variables and two outcome variables for analysis. 
\autoref{fig:annotation-window} shows the annotation labels and window construction.

\begin{itemize}
    \item \textbf{Predictor Variables}
    \begin{enumerate}
    \item \textbf{Concrete elaboration present:} 
    We set \(\mathrm{had\_elaboration\_in\_}AI_n = 1\) if the AI response at 
    \(n\) contained a concrete elaboration including an analogy, comparison-based 
    explanation, or example used to explain a concept.

    \item \textbf{Affective language present:} 
    We set \(\mathrm{had\_affective\_language\_in\_}AI_n = 1\) if the AI response 
    acknowledged student difficulty or used encouraging language.

    \item \textbf{Response length:} 
    We used the word count of the AI response at interaction \(n\) as a continuous 
    predictor. For regression analyses, response length was divided by 100 so that 
    model coefficients and odds ratios represented differences associated with each 
    additional 100 words. Response length was included alongside concrete elaboration 
    and affective language in the multivariable models.
\end{enumerate}

\item \textbf{Outcome Variables}
\begin{enumerate}
    \item \textbf{Re-confusion:} 
    We set \texttt{next\_confused} \(= 1\) when the student again showed confusion 
    in interaction \(n+1\) 
    (\(\mathrm{confusion\_in\_}U_{n+1} = 1\)), indicating that the student repeated 
    or rephrased the same confusion following the AI response at interaction \(n\). 
    Although both variables were derived from the same annotation label, 
    \(\mathrm{confusion\_in\_}U_n\) was used to identify the focal confusion windows, 
    whereas \(\mathrm{confusion\_in\_}U_{n+1}\) (\texttt{next\_confused}) served as 
    the subsequent-turn outcome.

    \item \textbf{Observed understanding:} 
    We constructed a composite next-turn indicator (`next\_understood`) to capture observable signs that the preceding confusion was no longer being pursued. The indicator was coded as 1 when the student either (a) explicitly expressed understanding of the preceding AI response (e.g., ``I see'' or 
    ``got it'') or (b) transitioned to a new topic without expressing continued confusion. This operationalization draws on the conversational principles of grounding and progressivity, whereby participants may proceed with an interaction without explicitly confirming comprehension when no further misunderstanding is raised \citep{ClarkBrennan1991grounding,albert2018repair}. In the context of naturalistic tutoring dialogues, a non-confused topic transition was therefore treated as an indirect behavioral indicator of conversational resolution rather than verified conceptual understanding. The measure captures immediate observable responses to AI feedback, not independently assessed learning gains. 
    
\end{enumerate}
\end{itemize}

\subsection{Human Validation of LLM annotations}\label{sec:human-validation}
To assess the reliability of the automated annotations,
we selected 150 previous--current interaction pairs,
including 100 randomly sampled pairs and 50 additional
pairs enriched for infrequent or potentially ambiguous
annotation categories. Two human annotators independently
coded the interactions using a shared codebook and the
relevant conversational context, without access to the
original LLM-generated labels. We assessed inter-annotator
agreement using percentage agreement and Cohen's
$\kappa$, and evaluated correspondence between the
original LLM labels and each human annotator using
Cohen's $\kappa$ and positive-class F1 scores.
The random and enriched subsets were identified
separately to account for the sampling design.
Detailed agreement statistics are provided in 
Appendix A (\autoref{tab:human-human-reliability} and \autoref{tab:human-llm-reliability}).

\subsection{Analytic Strategy}

We first used chi-square tests to describe unadjusted associations between each 
categorical feedback feature and the two subsequent-turn outcomes. Fisher's exact test was used when expected cell counts were below 5, and Cramér's \(V\) was reported as an effect-size measure for these unadjusted comparisons. Yates' continuity correction was applied to 2\(\times\)2 chi-square tests. All inferential tests used \(\alpha = .05\).

For the primary multivariable analyses, we fitted logistic Generalized Estimating 
Equation (GEE) models separately for next-turn understanding and re-confusion. 
Each model simultaneously included concrete elaboration, affective language, and 
continuous AI response length (scaled per 100 words) as predictors. GEE was used 
to account for repeated confusion windows contributed by the same user. Models 
used a binomial distribution with a logit link and an exchangeable working 
correlation structure, with observations clustered by user. Results are reported 
as odds ratios (ORs) with 95\% confidence intervals.

The primary GEE analyses included all 1,718 confusion windows from 152 users. 
Analyses were conducted in Python using \texttt{pandas}, \texttt{SciPy}, and 
\texttt{statsmodels}.

\section{Results}
\label{sec:results}

\subsection{Annotation Reliability and Validation}
\label{sec:annotation-reliability}

To evaluate the correspondence between the automated
annotations and human judgments, we compared the
original LLM-generated labels with independent human
annotations across 150 previous--current interaction
pairs. Human--LLM agreement varied by annotation
category. The strongest correspondence was observed for
Confusion ($\kappa = .922$--$.949$) and Understanding
($\kappa = .869$), whereas Follow-up and New Context
showed comparatively lower agreement
($\kappa = .651$--$.674$). Detailed category-level
agreement, human inter-annotator reliability, and
positive-class F1 scores are reported in
Appendix A.

\subsection{Descriptive Statistics}

Concrete elaboration and affective language rarely co-occurred; only 9 of 1,718 
windows (0.5\%) contained both features. \autoref{tab:descriptive-rates} summarizes 
the base rates for all predictors and outcomes. AI responses within the confusion 
windows had a mean length of 442.3 words (\(SD = 135.7\); median \(= 439\) words).

\begin{table}[t]
\centering
\caption{Predictor and outcome rates within confusion windows (\(N = 1{,}718\)).}
\label{tab:descriptive-rates}
\begin{tabular}{lrr}
\toprule
\textbf{Variables} & \textbf{\(N\)} & \textbf{\% of windows} \\
\midrule
\multicolumn{3}{l}{\textit{Predictors}} \\
Concrete elaboration present & 90 & 5.2\% \\
Affective language present & 198 & 11.5\% \\
\addlinespace
\multicolumn{3}{l}{\textit{Outcomes}} \\
Re-confusion (\texttt{next\_confused}) & 443 & 25.8\% \\
Understanding (\texttt{next\_understood}) & 201 & 11.7\% \\
\bottomrule
\end{tabular}
\end{table}

\subsection{Associations Between Micro-Level AI-Response Features and Subsequent-Turn Outcomes}

\begin{figure}[t]
    \centering
    \includegraphics[width=\linewidth]{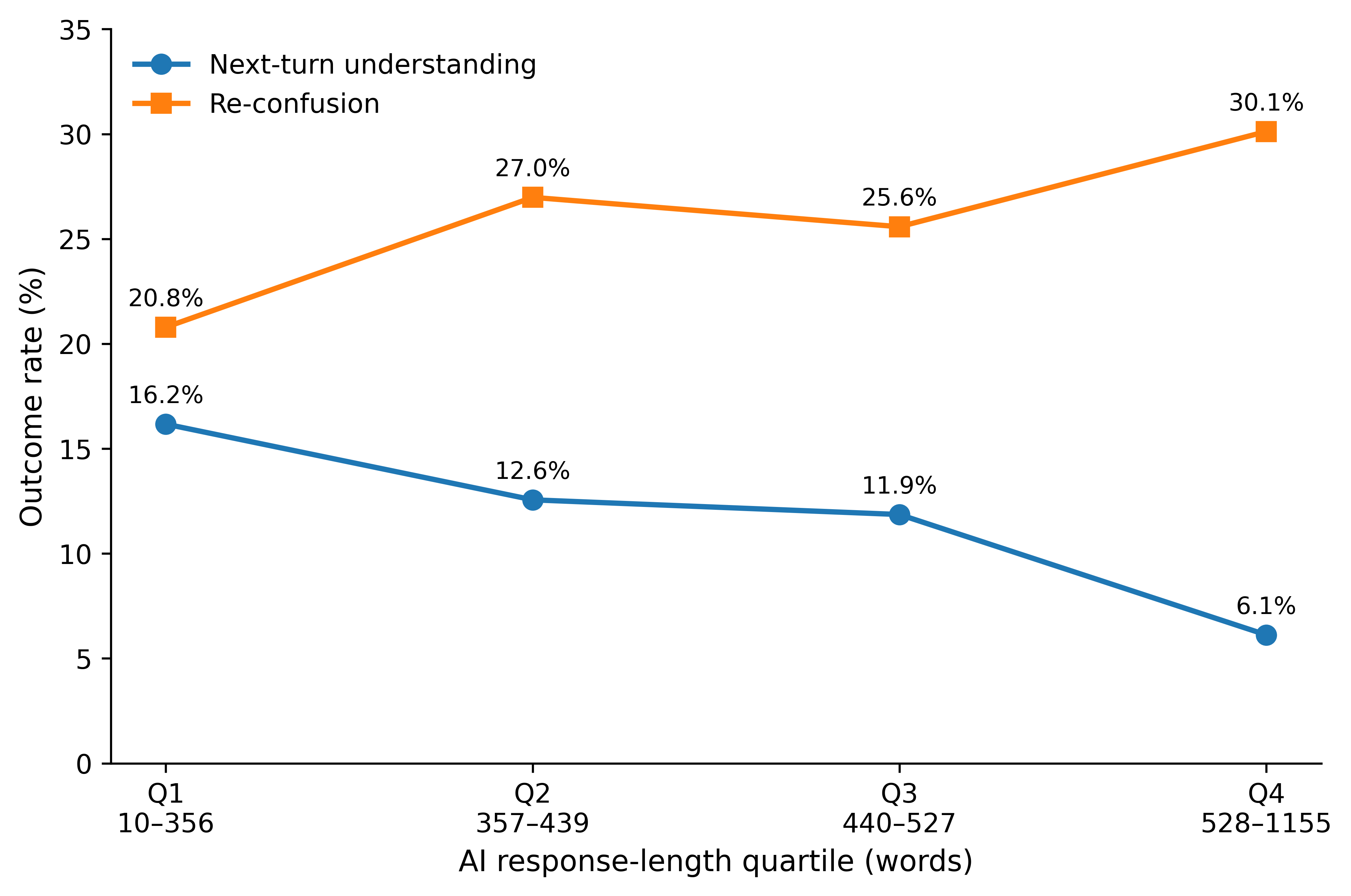}
    \caption{Descriptive rates of next-turn understanding and re-confusion across quartiles of AI response length.}
    \label{fig:length-quartiles}
\end{figure}

~\autoref{tab:chi-square} presents the unadjusted chi-square analyses for the 
categorical AI-response features. ~\autoref{fig:length-quartiles} descriptively 
displays outcome rates across response-length quartiles, while response length was 
retained as a continuous variable in all regression analyses. 
~\autoref{tab:gee-results} reports the GEE results accounting for within-user 
clustering.

\begin{table*}[h]
\centering
\caption{Chi-square tests of associations between AI-response features and 
subsequent-turn outcomes.}
\label{tab:chi-square}
\small
\begin{tabular}{llrrrr}
\toprule
\textbf{Predictor} & \textbf{Outcome} & \textbf{Present} & \textbf{Absent} & 
\(\boldsymbol{\chi^2 / p}\) & \textbf{\(V\)} \\
\midrule
Concrete Elaboration 
& Understanding 
& 21.1\% 
& 11.2\% 
& \(\chi^2 = 7.21,\ p = .007^{**}\) 
& .065 \\

& Re-confusion 
& 15.6\% 
& 26.4\% 
& \(\chi^2 = 4.69,\ p = .030^{*}\) 
& .052 \\
\addlinespace

Affective language 
& Understanding 
& 10.1\% 
& 11.9\% 
& \(\chi^2 = 0.39,\ p = .531\) 
& .015 \\

& Re-confusion 
& 31.3\% 
& 25.1\% 
& \(\chi^2 = 3.18,\ p = .075\) 
& .043 \\
\bottomrule
\end{tabular}

\vspace{2mm}
\begin{minipage}{0.96\linewidth}
\footnotesize
\textit{Note.} \(^{*}p < .05\), \(^{**}p < .01\). \(V\) = Cramér's \(V\).
\end{minipage}
\end{table*}

\begin{table*}[t]
\centering
\caption{GEE logistic regression predicting student outcomes following confused 
turns (\(N = 1{,}718\) windows, 152 users).}
\label{tab:gee-results}
\small
\begin{tabular}{lcccc}
\toprule
& \multicolumn{2}{c}{\textbf{Understanding}} 
& \multicolumn{2}{c}{\textbf{Re-confusion}} \\
\cmidrule(lr){2-3} \cmidrule(lr){4-5}
\textbf{Predictor} 
& \textbf{OR [95\% CI]} 
& \textbf{\(p\)} 
& \textbf{OR [95\% CI]} 
& \textbf{\(p\)} \\
\midrule

Concrete elaboration 
& 2.08 [1.23, 3.51] 
& .006 
& 0.52 [0.27, 1.00] 
& .050 \\

Affective language 
& 1.03 [0.63, 1.67] 
& .909 
& 1.16 [0.81, 1.68] 
& .417 \\

Response length (per 100 words) 
& 0.80 [0.72, 0.89] 
& \(< .001\) 
& 1.10 [1.03, 1.19] 
& .008 \\

\bottomrule
\end{tabular}

\vspace{2mm}
\begin{minipage}{0.96\linewidth}
\footnotesize
\textit{Note.} OR = odds ratio; CI = confidence interval. GEE models used a 
binomial distribution with a logit link and exchangeable working correlation 
structure, with observations clustered by user. Response length was scaled per 
100 words.
\end{minipage}
\end{table*}

\subsubsection{Concrete Elaboration and Subsequent-Turn Outcomes}

Concrete elaboration was associated with higher next-turn understanding in the 
unadjusted comparison (21.1\% vs.\ 11.2\%; 
\(\chi^2(1) = 7.21,\ p = .007,\ V = .065\)) and with lower re-confusion 
(15.6\% vs.\ 26.4\%; \(\chi^2(1) = 4.69,\ p = .030,\ V = .052\)). 
In the GEE model, concrete elaboration remained associated with higher odds of 
next-turn understanding (\(OR = 2.08\), 95\% CI [1.23, 3.51], \(p = .006\)) 
after accounting for affective language and response length. The estimated 
association with re-confusion was in the same direction 
(\(OR = 0.52\), 95\% CI [0.27, 1.00]), but did not reach the prespecified 
significance threshold (\(p = .050\)).

\subsubsection{Affective Language and Subsequent-Turn Outcomes}

Affective language was not significantly associated with either outcome in the 
unadjusted analyses. Understanding occurred in 10.1\% of windows with affective 
language and 11.9\% without it (\(p = .531\)), while re-confusion occurred in 
31.3\% and 25.1\% of windows, respectively (\(p = .075\)). Consistent with these 
unadjusted comparisons, the GEE models showed no detectable association between 
affective language and either next-turn understanding 
(\(OR = 1.03\), 95\% CI [0.63, 1.67], \(p = .909\)) or re-confusion 
(\(OR = 1.16\), 95\% CI [0.81, 1.68], \(p = .417\)).

\subsubsection{Response Length and Subsequent-Turn Outcomes}

Response length was associated with both subsequent-turn outcomes in the GEE 
models. Each additional 100 words in the AI response was associated with lower 
odds of next-turn understanding (\(OR = 0.80\), 95\% CI [0.72, 0.89], 
\(p < .001\)) and higher odds of re-confusion 
(\(OR = 1.10\), 95\% CI [1.03, 1.19], \(p = .008\)), after accounting for 
concrete elaboration and affective language. For descriptive visualization, 
outcome rates were also examined across response-length quartiles 
(~\autoref{fig:length-quartiles}). Understanding declined from 16.2\% in the 
shortest-response quartile to 6.1\% in the longest-response quartile, whereas 
re-confusion increased from 20.8\% to 30.1\%. Quartiles were used only for 
visualization; response length was modeled continuously in the inferential 
analyses.

\section{Discussion}
\label{sec:discussion}

\subsection{AI-Response Features Show Different Associations Across Consecutive Interactions}

This study examined whether three micro-level characteristics of AI responses---concrete elaboration, affective language, and response length---were associated with students' observed outcomes in the interaction immediately following a confused turn. The findings revealed distinct patterns across these response characteristics. Concrete elaboration was associated with higher odds of next-turn understanding, although its adjusted association with lower re-confusion narrowly missed the prespecified significance threshold. Affective language showed no detectable association with either outcome. In contrast, greater response length was associated with both lower odds of next-turn understanding and higher odds of re-confusion. Together, these findings demonstrate the value of examining AI-assisted feedback at the level of consecutive interactions, where associations between specific response characteristics and students' immediately subsequent behavior can be observed.

\subsection{RQ1: Concrete Elaboration and Subsequent-Turn Outcomes}

Concrete elaboration showed its clearest association with next-turn understanding. In the adjusted GEE model, AI responses containing an analogy, comparison, or example were associated with approximately twice the odds of observed understanding in the subsequent interaction compared with responses without concrete elaboration. The estimated association with re-confusion was also in the expected direction, with lower odds following concrete elaboration, but narrowly missed the prespecified significance threshold. Thus, the evidence is stronger for an association between concrete elaboration and observed understanding than for an association with re-confusion.

One possible interpretation is that concrete elaboration provides an explicit bridge between an unfamiliar concept and a more tangible representation at the point where a student has signalled difficulty. Analogies, comparisons, and worked examples may reduce the interpretive effort required to make sense of an explanation \citep{shutte2008feedback}. This interpretation is consistent with the observed association, but the present correlational design does not establish that concrete elaboration caused the subsequent change in student behavior.

Existing studies show that AI-generated analogies can support understanding, but that their benefit depends on subject matter, personalization, and how a teacher mediates their use \citep{CAO2024AnalogyNovice, Shao2025AnalogyLLM}. Separately, work on confusion in LLM tutoring dialogues has shown that tutors often respond to confusion with vague commentary rather than targeted scaffolding, but without isolating how specific response features relate to subsequent student behavior \citep{liu2025discerning,macina2023mathdial}. Our findings connect these strands by showing that naturally occurring concrete elaboration in LLM tutoring responses was associated with students' observed state in the immediately subsequent interaction. This suggests that associations previously discussed in more structured instructional settings can also be examined at a turn-to-turn level in naturalistic LLM tutoring conversations.

The association with re-confusion warrants particular caution. Although the unadjusted comparison indicated lower re-confusion when concrete elaboration was present, the adjusted GEE estimate narrowly missed the prespecified significance threshold. The direction of the estimate was consistent with the unadjusted pattern, but the confidence interval included the null value. Concrete elaboration was also relatively uncommon, occurring in only 5.2\% of confusion windows. Accordingly, the present evidence supports a clearer association with observed understanding than with re-confusion, and the latter should be examined in datasets containing more instances of concrete elaboration.

\subsection{RQ2: Affective Language and Subsequent-Turn Outcomes}

Affective language showed no detectable association with either next-turn outcome. Neither the unadjusted analyses nor the GEE models provided evidence that responses acknowledging difficulty or offering encouragement were associated with higher observed understanding or lower re-confusion. These findings should be interpreted as an absence of evidence for an association in this dataset and timescale, rather than evidence that affective language has no value in AI tutoring.

Prior work on affective AI feedback similarly suggests that emotionally supportive language may influence students' perceptions of feedback more consistently than behavioral or performance outcomes \citep{alsaiari2024emotionally, Ortega2024EmathyChatbot, Yin2025brainActivity}. Our analysis extends this question to the immediate interaction following a confused turn. One possible explanation is that brief affective phrases may serve a different function from explanatory content: acknowledging frustration or difficulty may shape the interaction experience without necessarily resolving the conceptual issue reflected in the next turn. The present analysis captures only this short timescale. Affective language may still contribute to outcomes such as longer-term engagement, persistence, or perceptions of the tutor, which cannot be evaluated using a single subsequent interaction.

\subsection{RQ3: Response Length and Subsequent-Turn Outcomes}

Response length showed associations with both subsequent-turn outcomes. After accounting for concrete elaboration and affective language, each additional 100 words in the AI response was associated with approximately 20\% lower odds of observed understanding and approximately 10\% higher odds of re-confusion. The descriptive quartile pattern was broadly consistent with these model estimates: understanding was lowest and re-confusion highest among responses in the longest quartile. Because word count was modeled continuously, this association does not depend on an arbitrary cut-point.

This pattern is consistent with prior work suggesting that students may find overly lengthy feedback difficult to process and often prefer focused responses \citep{alsaiari2024emotionally,siyan2024eden}. One possible interpretation is that students who have already expressed confusion may need to identify the most relevant information within the tutor's response, and additional text may increase processing demands or make the central explanation harder to identify. This interpretation is consistent with Cognitive Load Theory \citep{paas2003cognitive,sweller1988cognitive}, but cognitive load was not measured in the present study and therefore cannot be established as the mechanism underlying the observed association.

Read alongside the concrete elaboration finding, the response-length result highlights the distinction between the amount of information in a response and how that information is presented. Concrete elaboration was associated with higher observed understanding even after response length was included in the model, while greater response length itself was associated with lower understanding and higher re-confusion. These findings do not establish that shortening an AI response would improve student outcomes, because response length was not experimentally manipulated and may partly reflect characteristics of the student's original difficulty. They do, however, motivate controlled experiments that vary response length and concrete elaboration separately to determine whether concise, targeted explanations are more effective following student confusion.

\subsection{Limitations}

Several limitations should be considered. First, all annotations were generated using a single LLM-based pipeline. Agreement checks with human annotators produced encouraging results, but automated annotation may still introduce biases that differ from human judgment. Second, concrete elaboration and affective language were relatively rare within the confusion windows, which limited statistical precision, particularly for concrete elaboration estimates related to re-confusion. This low frequency is partly a consequence of using a secondary dataset: the underlying conversations were not collected with elaborate explanation or affective language in mind, and the tutor was not prompted to produce these features at any controlled rate. As a result, both features occur only incidentally, which limits how precisely their effects can be estimated. Third, the dataset represents one undergraduate AI course using one tutoring system. Whether these findings generalize to other subjects, educational levels, or AI models is unknown. 

Fourth, the composite understanding measure included non-confused topic transitions as indicators of conversational resolution. Although this approach captures students' observable progression beyond a previously expressed difficulty, a topic transition does not necessarily establish conceptual understanding. Students may move to another topic because they are satisfied with the explanation, choose to postpone the difficulty, or disengage from the original problem. Consequently, the outcome should be interpreted as an indicator of immediate conversational resolution rather than a direct measure of learning.

Fifth, AI response characteristics were likely related to properties of the preceding confused prompt. For example, more complex or persistent difficulties may have elicited longer or differently structured AI responses while also making subsequent resolution less likely. The models therefore account for the measured response features and repeated observations within users, but not for all potential differences in the nature or severity of the initial confusion.

Finally, this study is correlational. AI responses were not randomly assigned, and students who received a concrete elaboration may differ from those who did not in ways our variables do not capture. Experimental studies are needed to establish whether these response features cause changes in student outcomes, rather than reflecting association with them.

\subsection{Future Directions}

The low and uncontrolled base rates of concrete elaboration and affective language in this dataset point to a natural next step. Future studies could experimentally manipulate concrete elaboration, affective language, and response length rather than relying on features that occur incidentally in naturalistic conversations. In particular, independently varying concrete elaboration and response length would allow researchers to test whether concise responses containing targeted elaboration produce different subsequent outcomes from longer responses or responses without elaboration. Such designs could also incorporate direct measures of comprehension or task performance, allowing stronger causal tests of whether these response characteristics shape what students understand and do following confusion.

\section{Conclusion}\label{sec:conclusion}
This study examined whether three micro-level characteristics of AI tutor responses—concrete elaboration, affective language, and response length—were associated with students’ observed outcomes in the interaction immediately following a confused turn. By focusing on 1,718 confusion windows from naturalistic LLM tutoring conversations, the study provides a finer-grained view of AI-assisted feedback than analyses conducted only at the task or session level. Concrete elaboration was associated with higher odds of observed understanding in the subsequent interaction, although its adjusted association with lower re-confusion did not reach the prespecified significance threshold. Affective language showed no detectable association with either outcome, whereas greater response length was associated with lower odds of understanding and higher odds of re-confusion. Together, these findings show that response characteristics can exhibit distinct patterns of association at the level of consecutive interactions. They also motivate controlled experiments that manipulate concrete elaboration, affective language, and response length to determine whether these characteristics causally shape students’ responses following confusion. Future work should additionally examine whether these patterns generalize across subjects, student populations, and AI tutoring systems.




\appendix

\renewcommand{\thetable}{A\arabic{table}}
\setcounter{table}{0}

\bmsubsection*{Appendix A: Annotation Validation}
\label{app:annotation-validation}

We assessed the reliability of the annotation scheme
using two independent human annotators (H1 and H2) and
examined the correspondence between the human and
original LLM-generated annotations across six binary
categories. For analysis, blank annotation cells were
coded as 0, indicating absence of the corresponding
category. Follow-up and New Context were mutually
exclusive, whereas Confusion, Understanding, Concrete
Elaboration, and Affective Language were evaluated
independently.

~\autoref{tab:human-human-reliability} reports
inter-annotator agreement between H1 and H2. Agreement
was high across all six categories, ranging from
95.3\% to 100.0\%. Cohen's $\kappa$ ranged from .736
to 1.000, indicating substantial to perfect agreement
between the two human annotators.

\begin{table}[htbp]
\centering
\caption{Agreement between the two human annotators
($N = 150$ interaction pairs).}
\label{tab:human-human-reliability}
\small
\begin{tabular}{lrrr}
\toprule
\textbf{Category} &
\textbf{Agreement} &
\textbf{$\kappa$} &
\textbf{F1} \\
\midrule
Follow-up             & 95.3\%  & .856  & .971 \\
New Context           & 95.3\%  & .856  & .885 \\
Confusion             & 99.3\%  & .975  & .979 \\
Understanding         & 100.0\% & 1.000 & 1.000 \\
Concrete Elaboration  & 97.3\%  & .736  & .750 \\
Affective Language    & 98.0\%  & .869  & .880 \\
\bottomrule
\end{tabular}
\end{table}

~\autoref{tab:human-llm-reliability} reports pairwise
agreement between each human annotator and the original
LLM-generated annotations. Agreement was strongest for
Confusion and Understanding, with Cohen's $\kappa$
values ranging from .869 to .949. Agreement was also
substantial for Concrete Elaboration and Affective
Language. Follow-up and New Context showed comparatively
lower, although still substantial, agreement. These
comparisons indicate correspondence between the
LLM-generated annotations and independent human
judgments; however, they should not be interpreted as
accuracy estimates against an adjudicated human gold
standard.

\begin{table}[htbp]
\centering
\caption{Pairwise agreement between human annotators
and the original LLM-generated annotations
($N = 150$ interaction pairs).}
\label{tab:human-llm-reliability}
\small
\begin{tabular}{lrrrr}
\toprule
& \multicolumn{2}{c}{\textbf{H1--LLM}}
& \multicolumn{2}{c}{\textbf{H2--LLM}} \\
\cmidrule(lr){2-3}
\cmidrule(lr){4-5}
\textbf{Category} &
$\boldsymbol{\kappa}$ &
\textbf{F1} &
$\boldsymbol{\kappa}$ &
\textbf{F1} \\
\midrule
Follow-up             & .651 & .923 & .674 & .927 \\
New Context           & .651 & .727 & .674 & .746 \\
Confusion             & .922 & .933 & .949 & .957 \\
Understanding         & .869 & .880 & .869 & .880 \\
Concrete Elaboration  & .786 & .800 & .786 & .800 \\
Affective Language    & .706 & .727 & .782 & .800 \\
\bottomrule
\end{tabular}

\vspace{2mm}
\begin{minipage}{0.96\linewidth}
\footnotesize
\textit{Note.} H1 and H2 denote the two independent
human annotators. F1 is calculated for the positive
class (category present). Blank annotation cells were
coded as 0 (category absent). Human--LLM statistics
represent pairwise agreement and do not use adjudicated
human reference labels.
\end{minipage}
\end{table}

The validation set consisted of 150 interaction pairs:
100 randomly selected pairs and 50 additional pairs
selected to increase representation of less frequent
annotation categories. Because this enrichment changes
the prevalence of positive labels relative to the full
dataset, agreement statistics calculated from the
combined validation set should not be interpreted as
population-representative estimates.

\bmsection*{Data Availability Statement}
The StudyChat dataset is publicly available on Hugging Face (wmcnicho/StudyChat). Annotation prompts and analysis code are available from the corresponding author upon request.

\bmsection*{Ethics Statement}
This study is a secondary analysis of a publicly released, anonymized dataset. Ethical approval and informed consent were obtained by the original dataset creators.
\bmsubsection*{Financial Disclosure}

None reported.

\bmsubsection*{Conflicts of Interest}

The authors declare that they have no known competing financial interests or personal relationships that could have appeared to influence the work reported in this paper.


\bibliography{wileyNJD-Chicago}

\end{document}